\documentclass[conference]{IEEEtran}
\IEEEoverridecommandlockouts
\usepackage{cite}
\usepackage{float}
\usepackage{array}
\usepackage{multicol}
\usepackage{multirow}
\usepackage{booktabs}
\usepackage{graphicx}
\usepackage{forloop}
\usepackage{amsmath,amssymb,amsfonts}
\usepackage{algorithmic}
\usepackage{graphicx}
\usepackage{longtable}
\usepackage[colorlinks=true,linkcolor=black,urlcolor=black,breaklinks]{hyperref}
\usepackage{fancyhdr}
\usepackage{caption}

\usepackage{url}
\usepackage{textcomp}
\usepackage{xcolor}
\def\BibTeX{{\rm B\kern-.05em{\sc i\kern-.025em b}\kern-.08em
    T\kern-.1667em\lower.7ex\hbox{E}\kern-.125emX}}
\graphicspath{{Figures/}}

\makeatletter
\newcommand*\titleheader[1]{\gdef\@titleheader{#1}}
\AtBeginDocument{%
  \let\st@red@title\@title
  \def\@title{%
    \bgroup\normalfont\large\centering\@titleheader\par\egroup
    \vskip1.5em\st@red@title}
}
\makeatother

\title{Optimal MPPT Control of a Photovoltaic System Under Non-uniform Irradiation}
\titleheader{\small 2023 4th International Conference on Technological Advances in Electrical Engineering (ICTAEE)\\
University of 20 August 1955, Skikda, Algeria}
\author{\IEEEauthorblockN{Yehia Lalili}
\IEEEauthorblockA{\textit{\small Department of Electrical Engineering}\\
\textit{\small University of Skikda}\\
\small Skikda, Algeria \\
\href{mailto:y.lalili@univ-skikda.dz}{\texttt{y.lalili@univ-skikda.dz}}}
\and
\IEEEauthorblockN{Meriem Halimi}
\IEEEauthorblockA{\textit{\small Department of Automatic} \\
\textit{\small University of Jijel}\\
\small Jijel, Algeria \\
\href{mailto:halimi.meriem@gmail.com}{\texttt{halimi.meriem@gmail.com}}}
\and
\IEEEauthorblockN{Toufik Bouden}
\IEEEauthorblockA{\textit{\small Department of Automatic} \\
\textit{\small University of Jijel}\\
\small Jijel, Algeria \\
\href{mailto:bouden_toufik@yahoo.com}{\texttt{bouden\_toufik@yahoo.com}}}
}

\begin{document}

\maketitle

\begin{abstract}
Under Partial shading conditions (\textit{PSC}), traditional \textit{MPPT} methods such as, \textit{P\&O}, \textit{IncCon} and \textit{SMC}, cannot track down the Global \textit{MPP}. Thus, the energy conversion of the \textit{PV} modules will decrease. To overcome this drawback, two evolutionary algorithms; \textit{PSO} and \textit{CS}, were presented. Also, the study introduces a hybrid optimization method to enhance the overall performance of the \textit{PV} system under \textit{PSC}. \textit{MATLAB} simulations will be used in order to illustrate the efficiency of the proposed method.
\end{abstract}
\smallskip
\begin{IEEEkeywords}
maximum power point tracking, partial shading condition, particle swarm optimization, cuckoo search.
\end{IEEEkeywords}

\section{Introduction}
A large part of the energy consumed in the world relies on fossil fuels such as coal, gas, and oil. However, the resources for these energies will be exhausted in a few decades, and their exploitation produces toxic gases that are harmful to the environment. This has led to the development of other renewable and non-polluting sources of energy. Renewable energies are energies derived from inexhaustible natural resources such as water, heat from the earth, wind, and solar. In many countries, these energies have seen significant development, especially solar energy \cite{sarvi2022comprehensive}.\\

The power generated by photovoltaic (\textit{PV}) systems is largely dependent on weather conditions. Indeed, since the photovoltaic unit has nonlinear characteristics, the output power is considerably affected by changes in solar radiation, ambient temperature, and load. Therefore, a maximum power point tracking (\textit{MPPT}) technique designed to control the duty ratio of the \textit{DC/DC} converter is necessary to ensure optimal operation of the \textit{PV} chain in different operating conditions. Several studies have addressed the problem of finding the operating point that allows extracting the maximum energy from the \textit{PV} modules using various \textit{MPPT} methods (such as Perturbation and Observation (P\&O), Incremental Conductance (\textit{IncCon}), sliding mode, fuzzy logic, etc.) \cite{nihanth2020new}.\\

The problem of \textit{MPP} variations is further exacerbated by the fact that it is difficult to ensure that all \textit{PV} modules, which make up a \textit{PV} panel, are subjected to the same level of incident radiation, an unfavorable condition known as partial shading (\textit{PSC}). This condition results in a P-V curve with several peaks. The multiplicity of power peaks imposes expensive requirements on the \textit{MPPT} design due to the presence of local maximum points in addition to the global maximum power point (\textit{GMPP}). Conventional \textit{MPPT} techniques, such as \textit{IncCon} and P\&O, do not have the flexibility to differentiate between global and local points, which in turn leads to a reduction in the efficiency of the tracking system \cite{christopher2013comparative}.\\ 

In recent years, various modern techniques, which rely mainly on optimization techniques, have been explored to capture the \textit{GMPP} under \textit{PSC} \cite{moreira2019comparison}. Among these techniques, the Cuckoo Search (\textit{CS}) algorithm has been proposed \cite{ahmed2014maximum}. The \textit{CS} algorithm is inspired by the behavior of birds called "cuckoos" that perform brood parasitism by laying their eggs in the nests of other birds (called host birds). When searching for host bird nests, cuckoos choose directions or trajectories that can be modeled by certain mathematical functions, such as the \textit{Lévy} flight. To generate new cuckoos from existing ones, the \textit{CS} algorithm integrates the \textit{Lévy} flight as a means of generating new cuckoos from existing ones \cite{yang2009cuckoo}.\\

A comprehensive review of optimization techniques for renewable energy systems is presented in this article. The article covers two evolutionary algorithms, Particle Swarm Optimization (\textit{PSO}) and Cuckoo Search, as well as a hybrid optimization method designed to improve the overall performance of \textit{PV} systems under \textit{PSC}. The effectiveness of the proposed method will be demonstrated using \textit{MATLAB} simulations.\\ \\

This paper is organized to discuss various optimization techniques for renewable energy systems. In Section II, we focus on intelligent algorithms, specifically the Cuckoo search algorithm and the Particle swarm optimization. In Section III, we delve into the optimization of classical algorithms with a focus on Artificial Neural Networks (\textit{ANN}). Section IV presents simulation results for both the intelligent and optimized classical algorithms. Finally, in Section V, we draw conclusions based on our findings.

\section{Intelligent Algorithms}
\subsection{Cuckoo search (CS) algorithm}
The CS algorithm, originally proposed by \textit{Yang} and \textit{Deb} \cite{yang2009cuckoo}, is an algorithm based on the behavior of birds called ``cuckoo". Indeed, it has been observed that several species of cuckoo perform brood parasitism, that is, by laying their eggs in the nests of other birds (called host birds), previously observed. When searching for host bird nests, cuckoos choose directions or trajectories that can be modeled by certain mathematical functions. One of the most common models is \textit{Lévy} flight, which models the steps of the search for the cuckoo nest.\\

In order to generate new cuckoos from existing ones, Yang and Deb integrated the Lévy flight as follows:

\begin{equation}
x^{k+1}_{i} = x^{k}_{i}+\gamma\oplus \text{\textit{Lévy}}\left( \lambda\right),
\end{equation}
\begin{equation}
\gamma=\gamma_0\left( x^{k}_{best}-x^{k}_{i}\right),
\end{equation}
Where $x^{k}_{i}$ are samples (eggs), $i$ is the sample number, $k$  is the iteration number, $\gamma$  is the step size, and $\gamma_0$ is the initial step change.\\

A simplification of the Lévy distribution is given by \cite{ahmed2014maximum}:
\begin{equation}
\gamma_0 \left(x^{t}_{j} - x^{t}_{i}\right) \oplus \text{\textit{Lévy}}\left(\lambda\right) \approx K \times \left[\frac{u}{\left(\vert v\vert\right)^{\frac{1}{\beta}}}\right] \left( x^{k}_{best}-x^{k}_{i}\right),
\end{equation}
Where, $\beta=1.5$, $K$ is the \textit{Lévy} multiplier coefficient (chosen by the designer), while and are determined from the normal distribution curves:
\begin{equation}
u \approx N\left(0,\, \sigma^{2}_{u}\right),\quad v \approx N \left(0,\, \sigma^{2}_{v}\right).   
\end{equation}
Knowing that $\Gamma$ denotes the gamma function, then the variables $\sigma^{2}_{u}$  and $\sigma^{2}_{v}$ are defined by:
\begin{equation}
\sigma_{v}=1,\quad 
\sigma_{u} = \left[\frac{\Gamma \left(1+\beta\right) \times sin\left( \pi\times \frac{\beta}{2}\right) }{\Gamma\left(\frac{1+\beta}{2}\right)\times \beta \times 2^{\left(\frac{\beta - 1}{2}\right)}} \right]^{\frac{1}{\beta}}.
\end{equation}
To use the \textit{CS} algorithm to find the \textit{GMPP}, appropriate variables must be selected for the search.\\

First, there are the samples; in this case, they are defined as the voltage values of the \textit{GPV} (photovoltaic generator), i.e., $V_i=(i=1,2,\ldots,n)$ with $n$ being the total number of samples. Then, the step size, denoted as $\gamma$. 

The objective function $J$ is the value of the power generated by the \textit{PV} at the \textit{GMPP} ($P_{PM}$). Since depends on the \textit{PV} voltage, $J=P=f(V)$.\\

In the case of the \textit{GMPP} problem, and to ensure the search over the entire \textit{P-V} curve, the initial samples (particles) must be distributed over the entire voltage range. The number of samples $n$ is critical. A large $n$ increases the efficiency of the search (i.e., improves the chances of converging to a correct value), but requires a longer convergence time.\\

The \textit{CS} search mechanism in the case of \textit{PSC} is represented in Fig. 1. Three samples, designated by the variables $X$ (green), $Y$ (red) and $Z$ (yellow), are used. The superscript of the variable indicates the iteration number. The samples are initially distributed in various regions of the \textit{P-V} curve. In the first iteration, $Y^0$ is in the best position, therefore, $X^0$ and $Z^0$ are forced to leave their local positions and move towards $Y^0$.\\

However, in the second iteration, $Z^2$ reaches a better position than the others, so the other samples move towards it. It can be observed here that the true \textit{MPP} is slightly next to $Z^2$. Since the \textit{Lèvy} flight allows local samples to cross the best sample, $X$ and $Y$ cross $Z^2$ and also reach the \textit{GMPP} \cite{ahmed2014maximum}. 

\subsection{Particle swarm optimization (\textit{PSO})}
The \textit{PSO} algorithm is a bio-inspired approach based on animal social behavior, proposed in $1995$ by \textit{Eberhart} and \textit{Kennedy} \cite{eberhart1995new}. The algorithm starts with a random population of particles, where each particle/individual represents a possible solution. Each particle has its own velocity and position.\\

The main idea is to move particles in a way that they explore the search space for an optimal solution. The algorithm evaluates the performance of each particle position $x^{k}_{i}$ at each iteration $k$ through an objective function and changes its velocity towards its own best performance $x^{k}_{P\, best}$ and the best performance of the population $x^{k}_{G\, best}$.
\begin{figure}[b]
\centering
\includegraphics[scale=0.6]{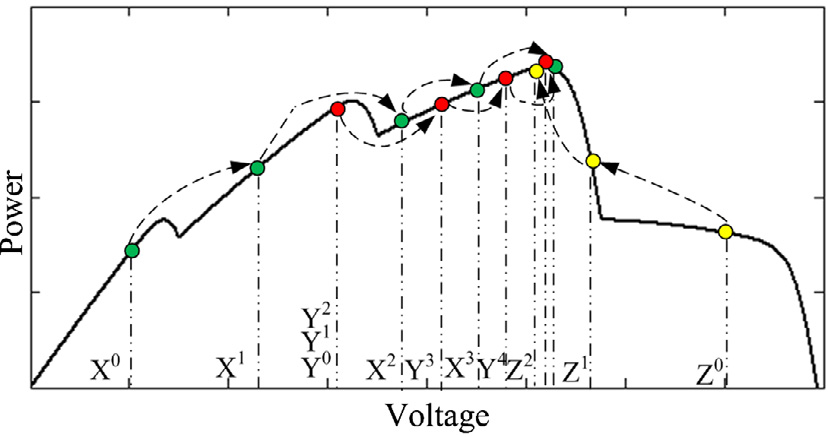}
\caption{\textit{GMPP} searching mechanism by \textit{CS} under \textit{PSC} \cite{ahmed2014maximum}.}
\label{fig:figure1}
\end{figure}
The velocity of each particle is calculated using (6), and the new position of the particles is determined by (7).
\begin{equation}
v^{k}_{i} = w v^{k-1}_{i}+r_1\alpha_1\left( x^{k}_{P\, best} - x^{k}_{i} \right) + r_2\alpha_2\left( x^{k}_{G\, best} - x^{k}_{i} \right)
\end{equation}
\begin{equation}
x^{k+1}_{i} = x^{k}_{i} + v^{k}_{i},
\end{equation}
Where $i\in \lbrace 1, \ldots, n\rbrace$  is the particle number, $k$ is the iteration number, $r_1$ and $r_2$ are random variables between $0$ and $1$, evaluated at each iteration following a uniform distribution, $\alpha_1$  and $\alpha_2$ are positive acceleration constants and $w$ is the inertial weight (usually chosen between $0.4$ and $0.9$).

To ensure that velocities and positions do not diverge, the parameters $\alpha_1$ and $\alpha_2$ must satisfy the following condition:
\begin{equation}
\alpha_1 + \alpha_2\leqslant 4.
\end{equation}
As with the \textit{CS} algorithm, and in order to perform the search over the entire \textit{P-V} curve, the initial samples must be distributed throughout the entire voltage range. The number of samples $n$ is also critical.\\

Extensive simulations indicate that $n=3$  is a good compromise and therefore it will be used in this work. In this case, the algorithm has a population of $3$ particles, where each particle can communicate with the others. Initially, the algorithm places the particles randomly in the search space (in the case of \textit{MPPT}, we are looking for the duty cycle $D$), and then the performance of each particle is evaluated by the output power of the \textit{PV} panel (objective function). Finally, the position of each particle is updated via (6) and (7), and the process is repeated. As the \textit{MPP} is approached, the velocity of each particle converges to zero.

\section{OPTIMIZATION OF CLASSICAL ALGORITHMS}
Consider a \textit{PV} system consisting of $3$ identical \textit{PV} modules connected in series, each with a bypass diode. By changing the irradiance on each module, the \textit{MPP} under various shading patterns can be generated, as shown in Fig. 2. It can be observed from this figure that the \textit{GMPP} is located in a specific region, namely $3$ zones of optimal voltage. 
\begin{figure}[bh]
\centering
\includegraphics[scale=0.75]{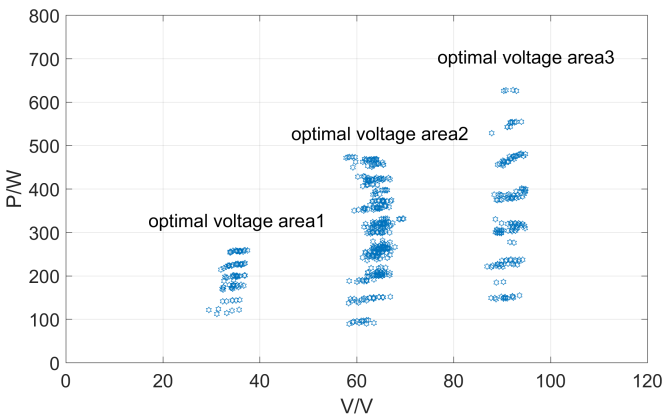}
\caption{\textit{GMPP} distribution of the \textit{PV} array under different \textit{PSC} \cite{zhang2019modified}.}
\label{fig:figure2}
\end{figure}
Based on this analysis, the basic idea of the method is to predict the optimal voltage zones, using artificial neural networks, namely $V_{min}$ and $V_{max}$, from the light intensity on different modules. Then, classical algorithm is applied to the zones to achieve the global maximum power point.
\begin{figure}[t]
\centering
\caption*{\small TABLE I. INPUT DATA AND OPTIMAL COORDINATES.}
\includegraphics[scale=0.8]{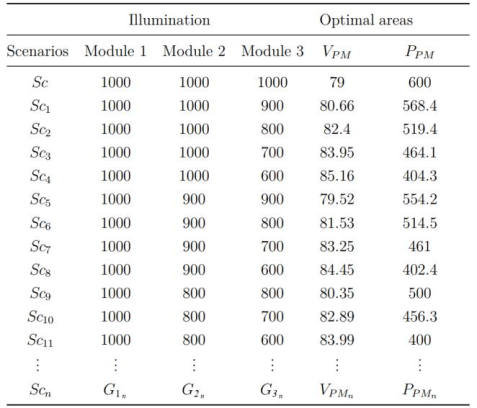}
\label{tab:table1}
\end{figure}
\subsection{Artificial neural network (\textit{ANN})}
Artificial neural networks (\textit{ANN}) are an interconnection of artificial neurons (nodes) that imitate a biological brain. They will be used to locate the optimal voltage zones of the \textit{GMPP} \cite{bouakkaz2020ann}.

One possible configuration of the \textit{ANN}, adapted to the \textit{MPPT}, consists of three layers, namely an input layer, a hidden layer, and an output layer, as shown in Fig. 3. In each layer, the number of nodes varies and is defined by the user. The illumination intensity $G$ on each module is introduced into the nodes of the input layer (the input data and optimal coordinates are represented in Table. I), then transferred to the hidden layer, which passes their output to the nodes of the output layer, getting the optimal voltage areas of \textit{GMPP}.\\

The training phase, which is an essential task to adjust the parameters of the \textit{ANN} structure, was carried out with the \textit{NNTOOL} tool of the \textit{MATLAB} software. This phase uses solar irradiation data. Fig. 4, presents the structure formed by the \textit{ANN} of the system simulated in \textit{MATLAB/Simulink} with $3$ input layers, $10$ hidden layers and $2$ output layers.
\begin{figure}[b]
\centering
\includegraphics[scale=0.65]{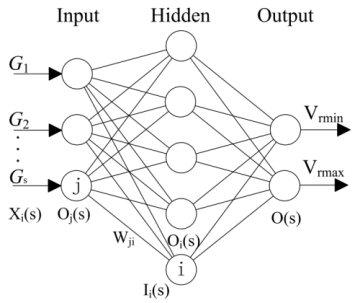}
\caption{Configuration of \textit{ANN} \cite{zhang2019modified}.}
\label{fig:figure3}
\end{figure}

\newpage
\section{SIMULATION RESULTS}
\subsection{Simulation of intelligent algorithms}
The simulation is performed to illustrate the capability of the \textit{PSO} and \textit{CS} algorithms in finding the \textit{GMPP} in the case of partial shading. The \textit{PV} panel is composed of three \textit{KG 200GT} modules connected in series (see Table. II) \cite{website}. \\

The modules receive the irradiance level of $600 \, W/m^2$, $800 \, W/m^2$, and $1000 \, W/m^2$, respectively. The power curve, illustrated in Fig. 6, shows that in this scenario, the \textit{PV} panel has three local \textit{MPPs} and one global \textit{MPP} corresponding to $P_{PM}=400 \, W$.
\begin{itemize}
\item[-] \textit{PSO} algorithm parameters are: $n=3$, $w=0.3$, and $\alpha_1=\alpha_2=1.2$. 

\item[-] \textit{CS} algorithm parameters are: $n=4$, $K=0.8$ and $\beta=1.5$. 

\item[-] Boost converter parameters are: $C_1=800 \, \mu F$, $C_2=850 \, \mu F$ and $L=0.0005 \, H$. 
\end{itemize}
The simulation results are shown in figs. 7, 8, 9, and 10.\\

\begin{figure}[H]
\centering
\caption*{\small TABLE II. SPECIFICATION OF THE \textit{PV} MODULE \textit{KC 200GT}.}
\includegraphics[scale=0.33]{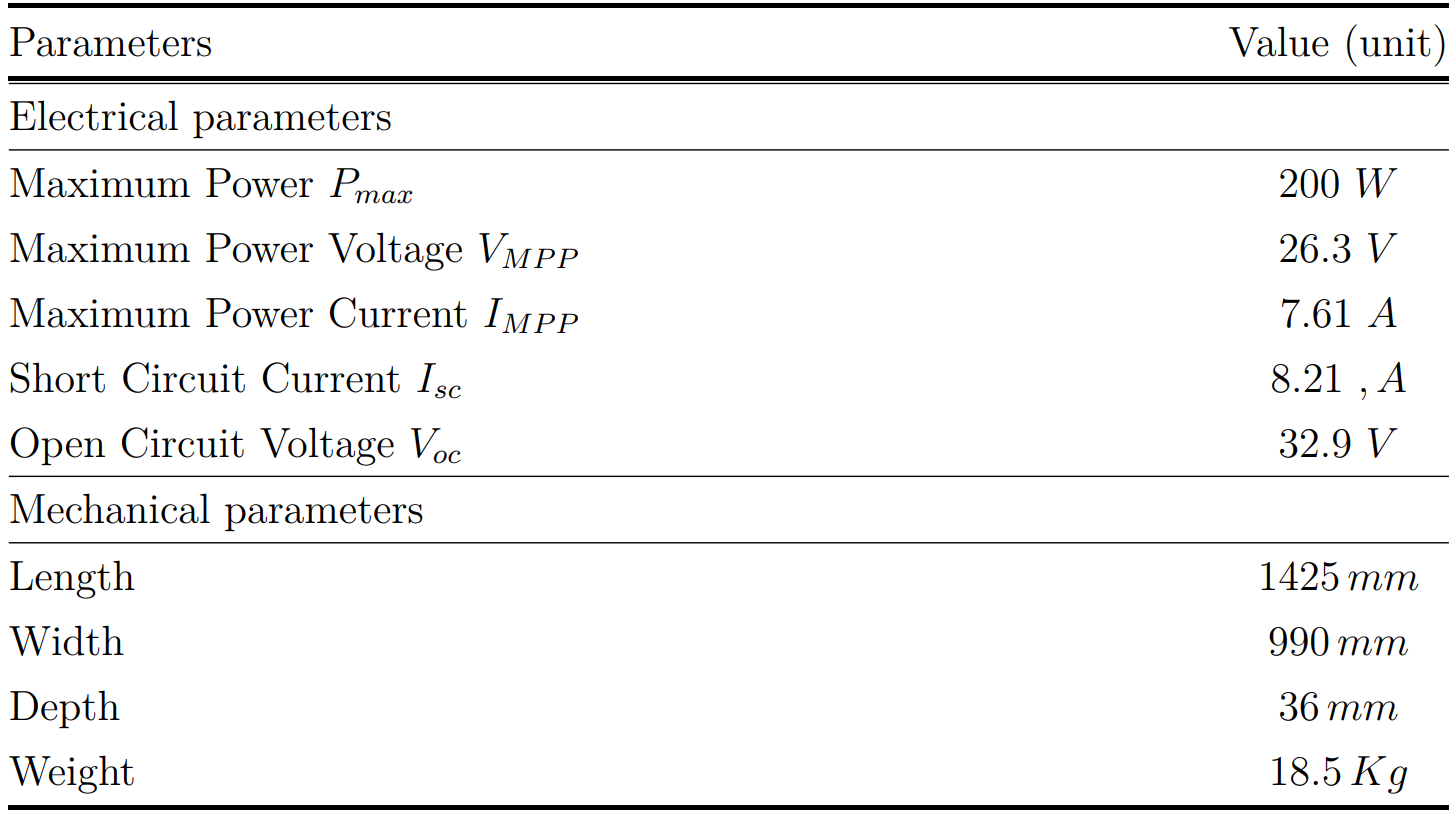}
\label{tab:table2}
\end{figure}
\begin{figure}[b]
\centering
\includegraphics[scale=0.6]{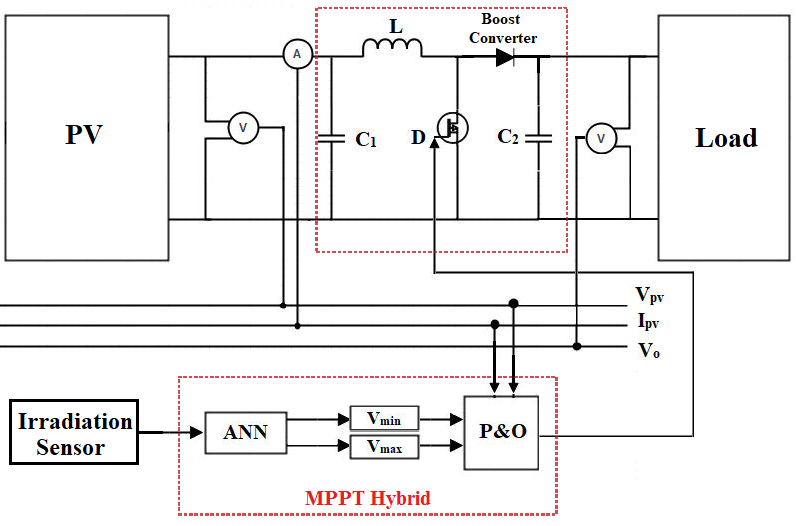}
\caption{Schematic diagram of the hybrid \textit{MPPT} control.}
\label{fig:figure5}
\end{figure}

According to the results obtained, it can be observed that:
\begin{itemize}
\item[-] \textit{MPPT} method based on \textit{CS} is faster and has a better response with lower oscillations than the \textit{PSO} method.
\item[-] Both the \textit{CS} and \textit{PSO} methods reach the \textit{GMPP} at $t = 0.7$ and $t = 1.2 \, s$ , respectively.
\item[-] The global maximum power is efficiently extracted via \textit{CS} in a short period of time compared to \textit{PSO}.
\end{itemize}

\subsection{Simulation of Optimized Classical Algorithms}
We used \textit{MATLAB/Simulink} to model and simulate the
\textit{PV} panels under different partial shading conditions (\textit{PSC}).
\begin{itemize}
\item[-] Boost converter parameters are: $C_1=800 \, \mu F$, $C_2=850 \, \mu F$ and $L=0.0005 \, H$.
\end{itemize}
The simulation results are shown in in figs. 11, 12, 13, and
14.\\

Simulation results indicate that:
\begin{itemize}
\item[-] The \textit{P\&O} method failed to reach the \textit{GMPP} and instead found a local \textit{PPM}, unlike the hybrid \textit{MPPT} method that was able to precisely extract the \textit{GMPP} with a high convergence speed. The \textit{ANN} allowed the \textit{P\&O} algorithm to limit its search to the \textit{GMPP's} vicinity by providing it with the values of $V_{min} = 79.99 \, V$ $and V_{max} = 88 \, V$.
\item[-] The proposed hybrid method exhibited fewer oscillations and reached the \textit{GMPP} more quickly ($t=0.18 \, s$) compared to the \textit{PSO} and \textit{CS} methods.
\end{itemize}

\begin{figure}[H]
\centering
\includegraphics[scale=0.9]{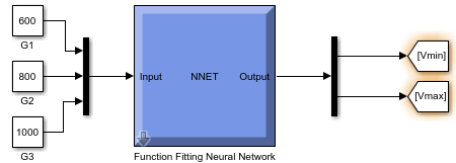}
\caption{ANN trained structure of the \textit{PV} system simulated in \textit{MATLAB/Simulink}.}
\label{fig:figure4}
\end{figure}
\begin{figure}[H]
\centering
\includegraphics[scale=0.42]{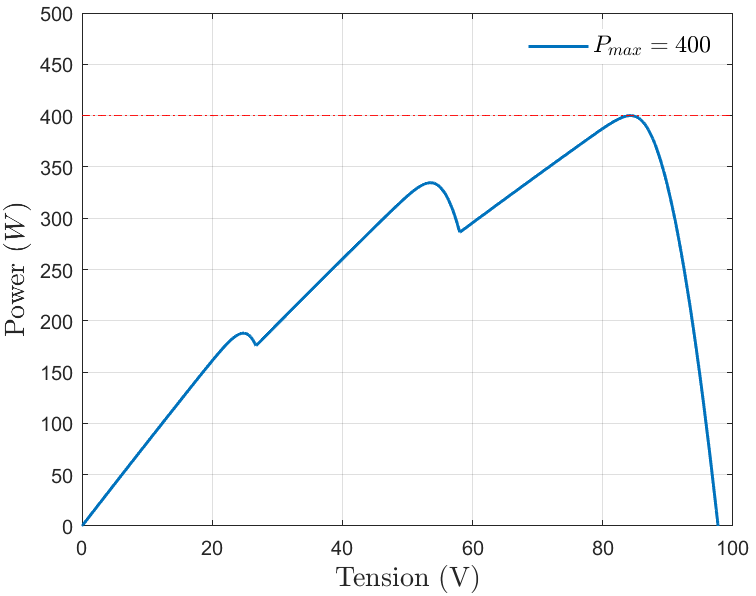}
\caption{P-V Curve under partial shading.}
\label{fig:figure6}
\end{figure}
\begin{figure}[H]
\centering
\includegraphics[scale=0.4]{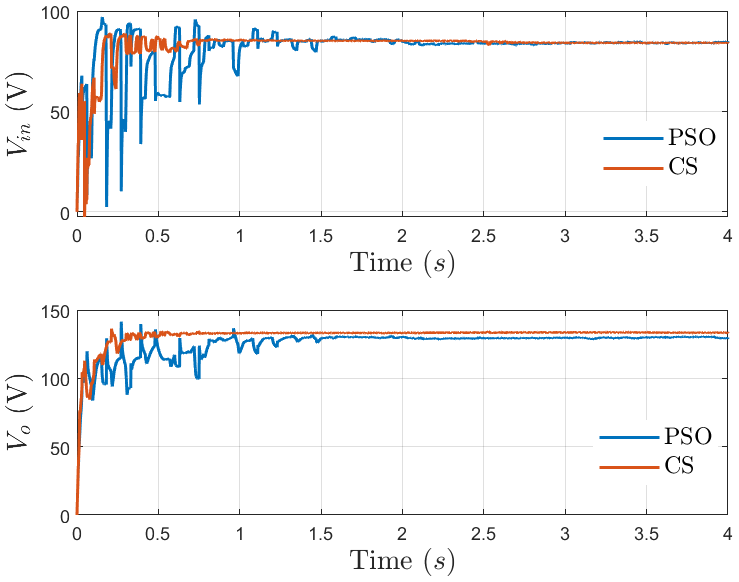}
\caption{Input and output voltages of the Boost converter with \textit{PSO} and \textit{CS}.}
\label{fig:figure7}
\end{figure}
\begin{figure}[H]
\centering
\includegraphics[scale=0.6]{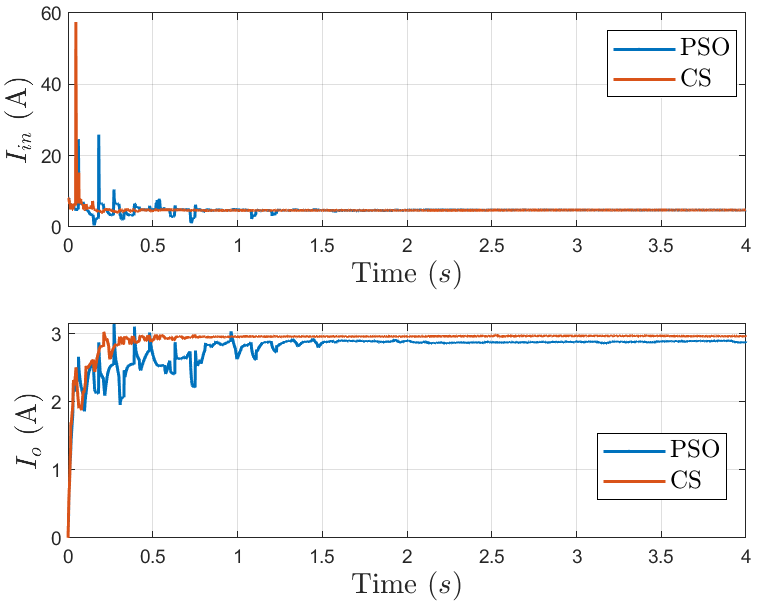}
\caption{Input and output currents of the Boost converter with \textit{PSO} and \textit{CS}.}
\label{fig:figure8}
\end{figure}
\begin{figure}[H]
\centering
\includegraphics[scale=0.6]{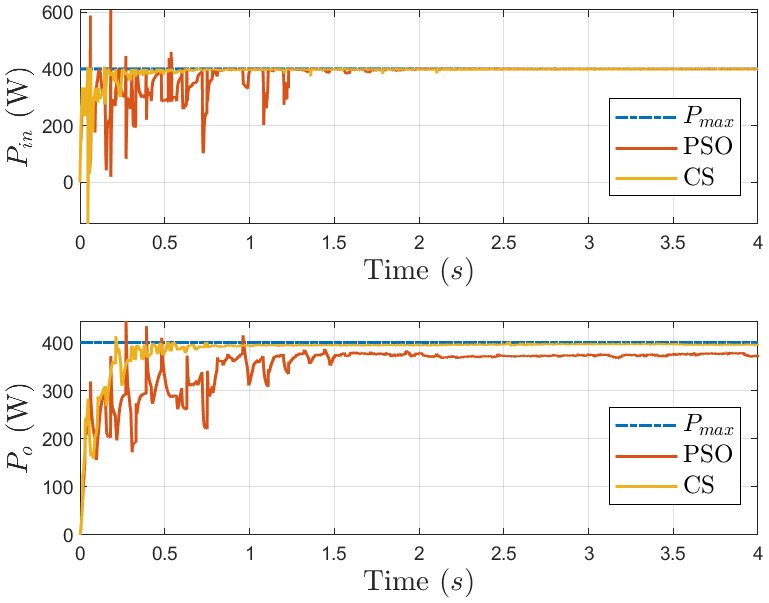}
\caption{GPV power and the transmitted power to the load with \textit{PSO} and \textit{CS}.}
\label{fig:figure9}
\end{figure}
\begin{figure}[H]
\centering
\includegraphics[scale=0.6]{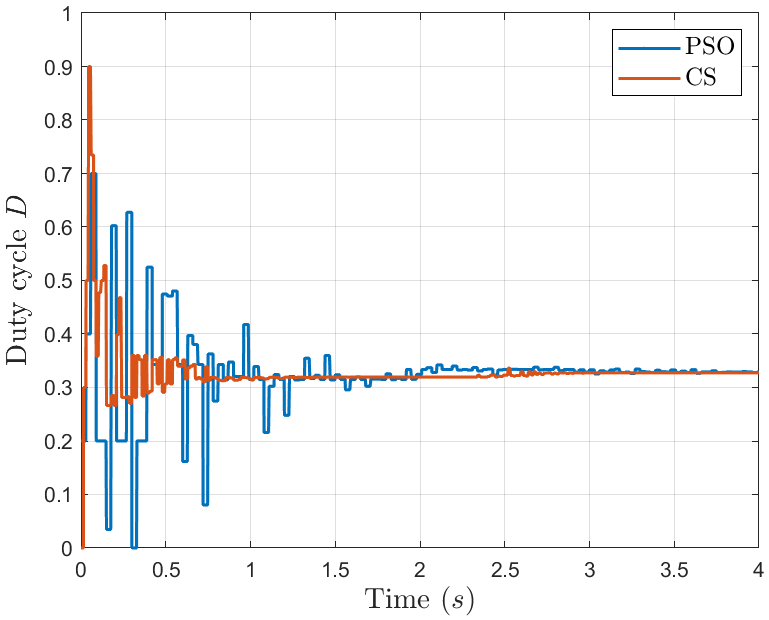}
\caption{Evolution of the Duty Cycle $D$ with \textit{PSO} and \textit{CS}.}
\label{fig:figure10}
\end{figure}

\begin{figure}[H]
\centering
\includegraphics[scale=0.4]{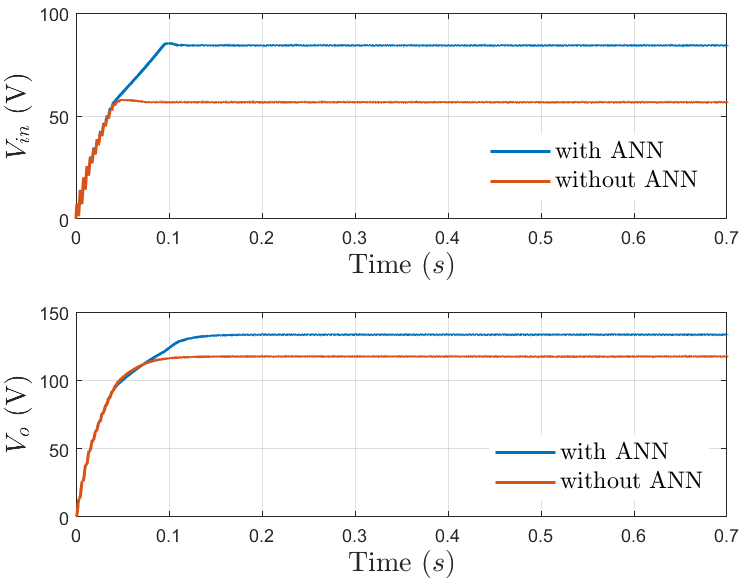}
\caption{Input and output voltages of the Boost converter with \textit{ANN}}
\label{fig:figure11}
\end{figure}

\begin{figure}[H]
\centering
\includegraphics[scale=0.4]{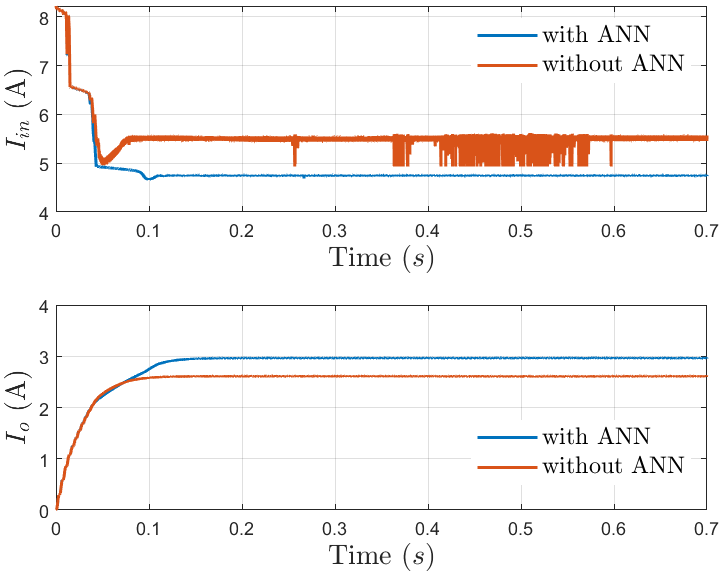}
\caption{Input and output currents of the Boost converter with \textit{ANN}}
\label{fig:figure12}
\end{figure}

\begin{figure}[t]
\centering
\includegraphics[scale=0.4]{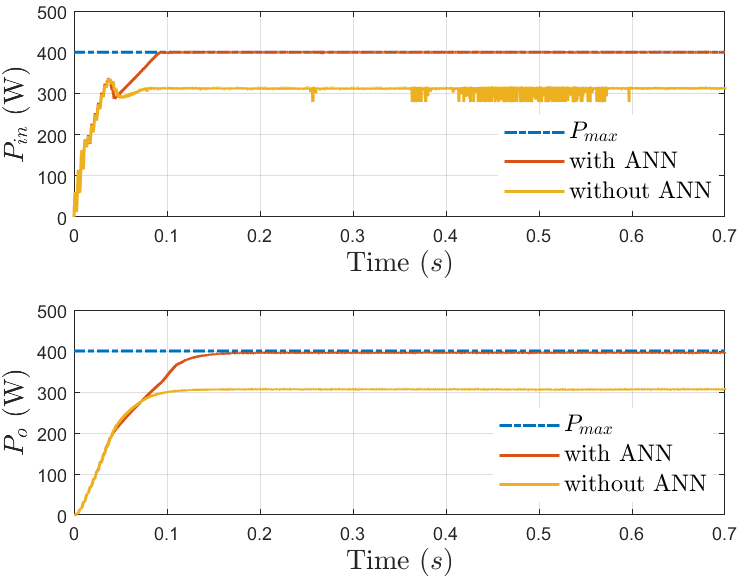}
\caption{\textit{GPV} power and the transmitted power to the load with \textit{ANN}}
\label{fig:figure13}
\end{figure}

\begin{figure}[t]
\centering
\includegraphics[scale=0.4]{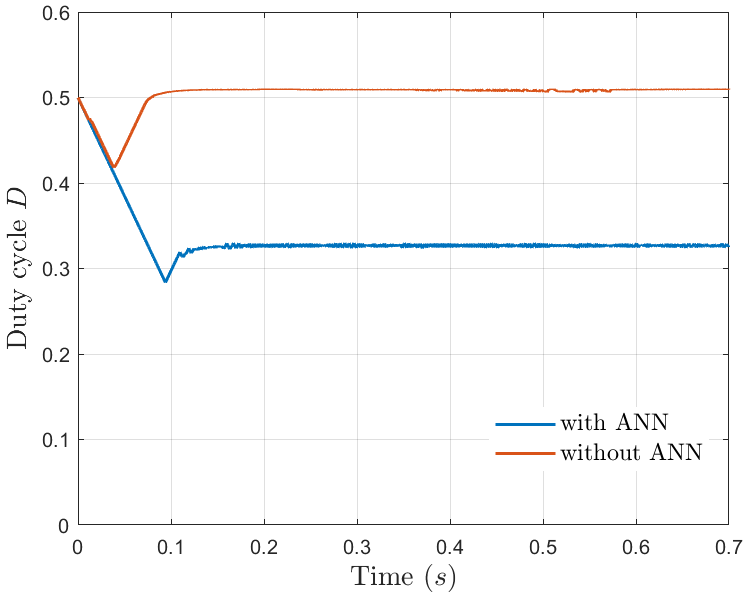}
\caption{Evolution of the Duty Cycle $D$ with \textit{ANN}.}
\label{fig:figure14}
\end{figure}

\newpage
\subsection{CONCLUSION}
In this paper, we have focused on the problem of nonuniform irradiation in photovoltaic systems, known as partial shading. An \textit{MPPT} control based on the use of intelligent algorithms \textit{CS} and \textit{PSO} was first studied. We found that the performance of the \textit{CS} algorithm is better than that of \textit{PSO} in terms of convergence speed towards the \textit{GMPP} and oscillations. Then, a hybrid \textit{MPPT} control, combining \textit{ANN} with the \textit{P\&O} algorithm, was studied. Simulation results showed that this method performs very well compared to intelligent algorithms. The hybrid method ensures a very fast convergence time and presents very low oscillations once the \textit{GMPP} is reached

\end{document}